\documentclass{article}

\makeatletter

\providecommand{\LyX}{L\kern-.1667em\lower.25em\hbox{Y}\kern-.125emX\@}

\makeatletter

\usepackage{amssymb}

\input{tcilatex}
\makeatother
\date{}
\makeatother

\begin{document}

\title{\textbf{Stochastic model for the dynamics of interacting Brownian particles}.}

\author{M. Mayorga\protect\protect\( ^{a}\protect \protect \), L. Romero-Salazar\protect\protect\( ^{a}\protect \protect \),
J.M. Rub\'{\i}\protect\protect\( ^{b}\protect \protect \)\\
 \protect\protect\( ^{a}\protect \protect \)\textit{Facultad de Ciencias}\\
 \textit{\ Universidad Aut\'{o}noma del Estado de M\'{e}xico,}\\
 \textit{Instituto Literario 100}\\
 \textit{\ 50000, Toluca, M\'{e}xico.}\\
 \protect\protect\( ^{b}\protect \protect \)\textit{Department de F\'{\i}sica
Fonamental,} \textit{Facultat de F\'{\i}sica,}\\
 \textit{\ Universitat de Barcelona,}\\
 \textit{\ Diagonal 647}\\
 \textit{08028, Barcelona, Spain.}}

\maketitle
\begin{abstract}
Using the scheme of mesoscopic nonequilibrium thermodynamics, we construct the
one- and two- particle Fokker-Planck equations for a system of interacting Brownian
particles. By means of these equations we derive the corresponding balance equations.
We obtain expressions for the heat flux and the pressure tensor which enable
one to describe the kinetic and potential energy interchange of the particles
with the heat bath. Through the momentum balance we analyze in particular the
diffusion regime to obtain the collective diffusion coefficient in terms of
the hydrodynamic and the effective forces acting on the Brownian particles. 
\end{abstract}

\section{Introduction}

\qquad Brownian effects are ubiquitous in many examples of soft condensed matter
physics \cite{cates} for which the system can be modelled as a set of interacting
degrees of freedom in contact with a heat bath. They play a very important role
when one infers macroscopic behaviors from the mesoscopic level of description,
a route utilized frequently in the study of complex systems. Many examples,
as colloidal suspensions \cite{russel3}, polymers \cite{Doi}, micelles \cite{cazabat},
etc., share those characteristics.

The dynamics at the mesoscopic level is governed by a set of Langevin processes
or equivalently by the corresponding \( N \)- particle Fokker-Planck equation.
One may arrive at the formulation of those equations by using different methods,
but the inherent complexity of the description of the dynamics of many interconnected
degrees of freedom makes it necessary to adopt simple schemes.

The method we use to analyze the dynamics of a system of interacting Brownian
particles, is based upon a mesoscopic approach proposed recently to obtain the
Fokker-Planck equation for the \( N \)-particle distribution function \cite{rubi}.
It applies the scheme of nonequilibrium thermodynamics to derive the kinetic
equation describing the evolution of the \( N \)-particle probability distribution
function \cite{mazur53,degroot}.

One then considers a system of \( N \) Brownian particles diluted in a solvent,
which acts as a thermal bath. The velocity of the particles are viewed as internal
thermodynamic variables. Such a description, is based on the existence of dissimilar
time scales for the Brownian particles and permits to proceed with an analysis
in the phase space of the macroparticles. The local equilibrium hypothesis is
introduced at the phase space level and from it one derives the entropy balance
equation. The resulting entropy production accounting for the irreversible processes
taking place in the phase space allows us to identify the fluxes and forces
in a similar manner as in the thermodynamic of the irreversible processes \cite{degroot}.

In the linear regime for the dependence of fluxes on the forces one obtains
the Fokker-Planck equations \cite{rubi,perezmadrid}. This theory has been applied
to the case of interacting Brownian particles under a temperature gradient,
and to the thermocapillary migration of Brownian droplets \cite{mayorga}. For
the case of Brownian motion under stationary flow, the mesoscopic approach permits
to obtain a Fokker-Planck equation which exhibits violation of the fluctuation-dissipation
theorem \cite{santamaria}, i.e., the diffusion coefficient does not obey the
Stokes-Einstein law. In addition, this formalism has been applied to obtain
kinetic equations for polymer dynamics \cite{rubi2}.

The above approach offers a mesoscopic framework for which simply by applying
the scheme of nonequilibrium thermodynamics in phase space it is possible to
derive kinetic equations of the Fokker-Planck or Smoluchowski types to analyze
the dynamics. Our goal in this paper is to apply that method to describe the
dynamics of interacting Brownian particles in the approximation of pair additive
interactions.

The paper has been organized in the following way. In section 2, we perform
a contraction of the description of the\ \( N \)-body problem to two tracer
particles, where their velocities are the internal degrees of freedom from the
thermodynamic point of view. Section 3 is devoted to obtain the Fokker-Planck
equations for the one- and two particles distribution functions. In section
4 we discuss the hydrodynamic regime, whereas in section 5 we analyze the diffusion
regime from the momentum balance equation. Finally, in section 6, we present
some concluding remarks.

\section{Conservation laws for the probability distribution functions}

\bigskip

\qquad We consider a set of \( N \) interacting Brownian particles of mass
\( m \)\textit{\ {}}suspended in a solvent which plays the role of a heat bath
at rest. We associate to each Brownian particle a phase coordinate vector \( \mathbf{x}_{i} \)
that denotes its position \( \mathbf{r}_{i} \), and its velocity \( \mathbf{u}_{i\text {}} \)
in phase space, i.e. \( \mathbf{x}_{i\text {}}=(\mathbf{r}_{i\text {}},\mathbf{u}_{i\text {}}) \).
Hence, the phase space occupied by the \( N \) particles is the one spanned
by the values of the vector \( \Gamma =(\mathbf{x}_{1},\mathbf{x}_{2},...\mathbf{x}_{N})\equiv (\mathbf{x}^{_{N}}). \)
The probability density for the \( N \) particles to be at point \( \Gamma  \)
at time \( t \) is denoted by \( P^{(N)}(\Gamma ,t) \) and satisfies the normalization
condition,

\begin{equation}
\int P^{(N)}(\Gamma ,t)d\Gamma =1.
\end{equation}

The entropy of the Brownian particles is given through the Gibbs entropy postulate,

\begin{equation}
S=-\, k\int P^{(N)}(\Gamma ,t)\ln \frac{P^{(N)}(\Gamma ,t)}{P_{eq.}^{(N)}(\Gamma )}d\Gamma +S_{eq.},
\end{equation}

\noindent where \( S_{eq.} \) is its equilibrium value, \( k \) is Boltzmann's
constant and \( P_{eq.}^{(N)} \) \( (\Gamma ) \) the \( N \)-particle equilibrium
probability density. In the framework of mesoscopic non-equilibrium thermodynamics
\cite{rubi,degroot}, the phase coordinates of the set of \( N \) Brownian
particles play the role of internal degrees of freedom. The Gibbs equation is
formulated as follows,

\begin{equation}
\delta S=-\, \frac{k}{T}\int \mu (\Gamma ,t)\delta P^{(N)}(\Gamma ,t)d\Gamma ,
\end{equation}
 where \( \mu (\Gamma ,t) \) is the non-equilibrium \( N \)-particle chemical
potential, \( T \) is the heat bath temperature which we assume constant and
\( \delta  \) stands for the exact total differential. The rate of variation
of the entropy is then given by \smallskip 
\begin{equation}
\frac{\partial S}{\partial t}=-\, \frac{k}{T}\int \mu (\Gamma ,t)\frac{\partial }{\partial t}P^{(N)}(\Gamma ,t)d\Gamma .
\end{equation}

Quite generally, we may assume that \( P^{(N)}(\Gamma ,t) \) satisfies the
continuity equation,

\begin{equation}
\frac{\partial P^{(N)}}{\partial t}+\sum _{i=1}^{N}\mathbf{u}_{i}\cdot \frac{\partial P^{(N)}}{\partial \mathbf{r}_{i}}-m^{-1}\sum _{i,j=1}^{N}\frac{\partial \phi _{ij}}{\partial \mathbf{r}_{i}}\cdot \frac{\partial P^{(N)}}{\partial \mathbf{u}_{i}}=-\sum _{i=1}^{N}\frac{\partial }{\partial \mathbf{u}_{i}}\cdot \mathbf{J}_{u_{i}}^{(N)},
\end{equation}

\noindent where \( \phi _{ij} \) represents the direct interaction potential
between particles. The energy and momentum dissipative interchange between particles
and the solvent is taken into account through the fluxes \( \mathbf{J}_{u_{i}}^{(N)}=\mathbf{J}_{u_{i}}^{(N)}(\mathbf{x}^{N}) \).

From the \( N \)-particle continuity equation (5), after integrating over \( N \)-\( s \)
phase coordinates, we obtain the reduced continuity equations,

\begin{equation}
\frac{\partial P^{(s)}}{\partial t}+\sum _{i=1}^{s}\mathbf{u}_{i}\cdot \frac{\partial P^{(s)}}{\partial \mathbf{r}_{i}}-m^{-1}\int \sum _{i,j=1}^{s}\frac{\partial \phi _{ij}}{\partial \mathbf{r}_{i}}\cdot \frac{\partial P^{(N)}}{\partial \mathbf{u}_{i}}d\mathbf{x}_{s+1}...d\mathbf{x}_{N}=-\sum _{i=1}^{s}\frac{\partial }{\partial \mathbf{u}_{i}}\cdot \mathbf{J}_{u_{i}}^{(s)},
\end{equation}

\noindent with \( P^{(s)}=P^{(s)}(\mathbf{x}^{s},t) \)
\begin{equation}
P^{(s)}(\mathbf{x}_{1},\mathbf{x}_{2},...\mathbf{x}_{s},t)=\frac{N!}{(N-s)!}\int ^{(N-s)}P^{(N)}d\mathbf{x}_{s+1}...d\mathbf{x}_{N}\text {,}
\end{equation}
 the reduced distribution function and

\begin{equation}
\mathbf{J}_{u_{i}}^{(s)}=\int \mathbf{J}_{u_{i}}^{(N)}d\mathbf{x}_{s+1}...d\mathbf{x}_{N},
\end{equation}
 the reduced fluxes in phase space.

\smallskip In particular, we are interested in the case of the one- and two-particle
distribution functions, hence from the above equation we can obtain, the continuity
equations,

\smallskip 
\begin{equation}
\frac{\partial }{\partial t}P^{(1)}+\mathbf{u}_{1}\cdot \frac{\partial }{\partial \mathbf{r}_{1}}P^{(1)}-m^{-1}\int \frac{\partial \phi _{12}}{\partial \mathbf{r}_{1}}\cdot \frac{\partial P^{(2)}}{\partial \mathbf{u}_{1}}d\mathbf{x}_{2}=-\frac{\partial }{\partial \mathbf{u}_{1}}\cdot \mathbf{J}_{u_{1}}^{(1)}
\end{equation}
 and 
\begin{equation}
\frac{\partial }{\partial t}P^{(2)}+\sum _{i=1}^{2}\mathbf{u}_{i}\cdot \frac{\partial P^{(2)}}{\partial \mathbf{r}_{i}}-m^{-1}\sum _{i,j=1}^{2}\frac{\partial \phi _{ij}}{\partial \mathbf{r}_{i}}\cdot \frac{\partial P^{(2)}}{\partial \mathbf{u}_{i}}=-\sum _{i=1}^{2}\frac{\partial }{\partial \mathbf{u}_{i}}\cdot \mathbf{J}_{u_{i}}^{(2)}.
\end{equation}

\noindent These equations present the usual drift terms and conservative interaction
contributions on the left hand side. In eq. (9) the third term expresses the
mean force over a Brownian particle due to the presence of other particles.
On the other hand, eq. (10) includes the effect of the direct forces felt by
a pair of particles. The right hand side of each equation accounts for the dissipative
interaction of the particles with the solvent through the corresponding currents.
At this stage, we have provided the continuity equations for the reduced probability
densities \( P^{(1)} \) and \( P^{(2)}, \) where the fluxes \( \mathbf{J}_{u_{i}}^{(i)} \)
appear as unknown functions. One of the features of the mesoscopic nonequilibrium
thermodynamics approach presented here, is to find out the expressions of \( \mathbf{J}_{u_{1}}^{(1)} \)
and \( \mathbf{J}_{u_{i}}^{(2)} \) , which will be obtained from the entropy
production in the next section.

\bigskip

\section{Fokker-Planck equations for the one- and two- particle distribution functions}

\smallskip \qquad The simplest model of N-interacting particles can be implemented
by only considering interactions between pairs. This approximation implies a
substantial simplification which in terms of our mesoscopic description concerns
the analysis of the dynamics through the one and two-particle distribution functions.
In this section we will derive the kinetic equations of the Fokker-Planck type,
accounting for the evolution of those distribution functions.

To proceed, we factorize the N-particle distribution function \( P^{(N)}(\mathbf{x}^{N},t) \)
as follows

\begin{equation}
P^{(N)}(\mathbf{x}^{N},t)=P^{(1)}(\mathbf{x}_{1},t)P^{(1)}(\mathbf{x}_{2},t)...P^{(1)}(\mathbf{x}_{N},t)g^{(N)}\left( \mathbf{x}^{N},t\right) ,
\end{equation}
 where \( P^{(1)}\left( \mathbf{x}_{i},t\right)  \) represents the reduced
distribution function for \( s=1; \) moreover, \( g^{(N)}\left( \mathbf{x}^{N},t\right)  \)
is the \( N \)-particle dynamic correlation function. In a similar manner,
as for the equilibrium situation {[}13-15{]}, we propose a factorization for\ the
correlation function \( g^{(N)}\left( \mathbf{x}^{N},t\right) = \) \( g^{(N)}(\mathbf{x}_{1},\mathbf{x}_{2},...,\mathbf{x}_{N},t) \)
as follows,\bigskip 
\begin{eqnarray}
g^{(N)} & = & g^{(2)}\left( \mathbf{x}_{1},\mathbf{x}_{2},t\right) ...g^{(2)}\left( \mathbf{x}_{N-1},\mathbf{x}_{N},t\right) \delta g^{(3)}\left( \mathbf{x}_{1},\mathbf{x}_{2},\mathbf{x}_{3},t\right) ...\nonumber \\
 &  & \delta g^{(3)}\left( \mathbf{x}_{N-2},\mathbf{x}_{N-1},\mathbf{x}_{N},t\right) ...\delta g^{(N)}\left( \mathbf{x}_{1},...,\mathbf{x}_{N},t\right) .
\end{eqnarray}
 Here, \( \delta g^{(3)}\left( \mathbf{x}_{1},\mathbf{x}_{2},\mathbf{x}_{3},t\right)  \)
is defined through the relation

\begin{equation}
g^{(3)}\left( \mathbf{x}_{1},\mathbf{x}_{2},\mathbf{x}_{3},t\right) =g^{(2)}\left( \mathbf{x}_{1},\mathbf{x}_{2},t\right) g^{(2)}\left( \mathbf{x}_{1},\mathbf{x}_{3},t\right) g^{(2)}\left( \mathbf{x}_{2},\mathbf{x}_{3},t\right) \delta g^{(3)}\left( \mathbf{x}_{1},\mathbf{x}_{2},\mathbf{x}_{3},t\right) ,
\end{equation}
 such that when \( \delta g^{(3)}\left( \mathbf{x}_{1},\mathbf{x}_{2},\mathbf{x}_{3},t\right) =1 \),
it reduces to the so-called Kirkwood's superposition approximation, and \( g^{(4)} \)
is defined similarly, through \( \delta g^{(N)} \). In the above factorization
for \( g^{(N)}, \) we have \( \left( 1/2\right) N\left( N-1\right)  \) pairs,
\( \left( 1/3!\right) N\left( N-1\right)  \) \( \left( N-2\right)  \) triplets,
etc. so that

\noindent 
\begin{eqnarray}
\ln P^{(N)} & = & \sum _{i=1}^{N}\ln \left( P^{(1)}\left( \mathbf{x}_{i},t\right) \right) +\ln g^{(N)}\left( \mathbf{x}^{N},t\right) \nonumber \\
 & = & N\ln P^{(1)}\left( \mathbf{x}_{1},t\right) +\frac{N}{2}\left( N-1\right) \ln g^{(2)}\left( \mathbf{x}_{1},\mathbf{x}_{2},t\right) +\nonumber \\
 &  & \frac{N}{3!}\left( N-1\right) \left( N-2\right) \ln \delta g^{(3)}\left( \mathbf{x}_{1},\mathbf{x}_{2},\mathbf{x}_{3},t\right) +...
\end{eqnarray}
 Similary, for the equilibrium case {[}13-15{]}, we have 
\begin{equation}
\ln P_{eq}^{(N)}\left( \mathbf{x}^{N}\right) =N\ln P_{eq}^{(1)}\left( \mathbf{x}_{1}\right) +\frac{N}{2}\left( N-1\right) \ln g_{eq}^{(2)}\left( \mathbf{r}_{1},\mathbf{r}_{2}\right) +...\text {}.
\end{equation}
 \qquad \qquad \smallskip Substitution of these \ two last equations in (2),
yields the expression of the Gibbs entropy postulate,

\begin{eqnarray}
S & = & -Nk\int P^{(1)}\left( \mathbf{x}_{1},t\right) \ln \frac{P^{(1)}\left( \mathbf{x}_{1},t\right) }{P_{eq}^{(1)}\left( \mathbf{x}_{1}\right) }d\mathbf{x}_{1}\nonumber \\
 &  & -\frac{N\left( N-1\right) }{2}k\int P^{(2)}\left( \mathbf{x}_{1},\mathbf{x}_{2},t\right) \ln \left( \frac{\, \, g^{(2)}\left( \mathbf{x}_{1},\mathbf{x}_{2},t\right) }{g_{eq}^{(2)}\left( \mathbf{r}_{1},\mathbf{r}_{2}\right) }\right) d\mathbf{x}_{1}d\mathbf{x}_{2}+\nonumber \\
 &  & ...+S_{eq}.
\end{eqnarray}

\noindent It is worth mentioning that a similar expression, for the non-equilibrium
entropy, has been obtained for a kinetic description of a dense gas \cite{romero}
and its equilibrium counterpart has also been used for a calculation of the
thermodynamic entropy \cite{wallace,evans}. In addition, this statistical entropy
expression has been recently reformulated \ in terms of the potentials of mean
force between particles \cite{pouskari}, and their graphical representation
has been presented for the excess entropy \cite{bednorz}.

\smallskip

\smallskip Similarly to the \( N \)-particle case, we formulate the Gibbs equation
as follows,

\begin{eqnarray}
\delta S & = & -\frac{Nm}{T}\int \mu ^{(1)}(\mathbf{x}_{1},t)\delta P^{(1)}d\mathbf{x}_{1}\nonumber \\
 &  & -\frac{N\left( N-1\right) m^{2}}{2T}\int \mu ^{(2)}(\mathbf{x}_{1,}\mathbf{x}_{2},t)\delta P^{(2)}d\mathbf{x}_{2}d\mathbf{x}_{1},
\end{eqnarray}
 where \( \mu ^{(1)}+\mu ^{(2)} \) represents the non-equilibrium chemical
potential per unit mass of the suspended particles. Their general forms are
given by,

\begin{equation}
\mu ^{(1)}(\mathbf{x}_{1,}t)=\frac{kT}{m}\ln P^{(1)}(\mathbf{x}_{1,}t)+\Psi ^{(1)}(\mathbf{x}_{1,}t)
\end{equation}
 and

\begin{equation}
\mu ^{(2)}(\mathbf{x}_{1,}\mathbf{x}_{2},t)=\frac{kT}{m}\ln g^{\left( 2\right) }(\mathbf{x}_{1,}\mathbf{x}_{2},t)+\Psi ^{(2)}(\mathbf{x}_{1,}\mathbf{x}_{2},t),
\end{equation}
 where \( \Psi ^{(1)} \) and \( \Psi ^{(2)} \) are unknown potential functions.

At equilibrium, the distribution functions must satisfy the next expressions,

\begin{equation}
P_{eq.}^{(1)}\left( u_{1}\right) =\exp \frac{m}{kT_{eq}}\left( \mu _{B_{eq}}^{id}-\frac{u_{1}^{2}}{2}\right) ,
\end{equation}
 where we have defined the distribution function in a reference frame moving
with the mean velocity of Brownian particles and

\begin{equation}
g_{eq}^{(2)}\left( \mathbf{r}_{1},\mathbf{r}_{2}\right) =\exp \frac{m}{kT_{eq}}\left( \mu _{B_{eq}}^{exc}\left( \mathbf{r}_{1},\mathbf{r}_{2}\right) -\phi _{12}\left( \mathbf{r}_{1},\mathbf{r}_{2}\right) \right) .
\end{equation}
 Here \( \mu _{B_{eq}}^{id} \) and \( \mu _{B_{eq}}^{exc}\left( \mathbf{r}_{1},\mathbf{r}_{2}\right)  \)
correspond to the ideal and excess parts of the chemical potential of the Brownian
particles at equilibrium with the solvent. On the other hand, according to the
above definition of the pair correlation function we can identify the effective
pair interaction potential, namely, \( \phi _{12}^{eff}\left( \mathbf{r}_{1},\mathbf{r}_{2}\right) =\phi _{12}\left( \mathbf{r}_{1},\mathbf{r}_{2}\right) -\mu _{B_{eq}}^{exc}\left( \mathbf{r}_{1},\mathbf{r}_{2}\right) , \)
where \( -\mu _{B_{eq}}^{exc} \) stands for the indirect part of the work,
corresponding to the free energy change, to transport the Brownian particles
at a certain separation, accounting for the static interaction of the particles
with the solvent. For the particular case when the effective interaction \( \phi _{12}^{eff}\left( \mathbf{r}_{1},\mathbf{r}_{2}\right)  \)
has spherical symmetry, \( g_{eq}^{(2)} \) only depends on the distance separation
modulus of the two particles \( r_{12}=\mid \mathbf{r}_{12}\mid =\mid \mathbf{r}_{2}-\mathbf{r}_{1}\mid  \),
i.e. \( g_{eq}^{(2)}\left( \mid \mathbf{r}_{2}-\mathbf{r}_{1}\mid \right)  \)
\cite{hansen}.

If we now substitute (20) and (21) in (18) and (19), such that in this limit
case \( \mu ^{(1)}= \) \( \mu _{B_{eq}}^{id} \) and \( \mu ^{(2)}=\mu _{B_{eq}}^{exc}, \)
and\ using the definition of the total chemical potential at equilibrium \( \mu _{B_{eq}}^{tot}= \)
\( \mu _{B_{eq}}^{id}+ \) \( \mu _{B_{eq}}^{exc}, \) one obtains the expressions

\begin{equation}
\Psi ^{(1)}=\frac{u_{1}^{2}}{2},
\end{equation}
 and

\begin{equation}
\Psi ^{(2)}=\phi _{12}.
\end{equation}
 With this identification, we can rewrite expressions (18) and (19) as follows,

\begin{equation}
\mu ^{(1)}=\frac{kT}{m}\ln \left( \frac{f^{(1)}}{f_{eq}^{(1)}}\right) +\mu _{B_{eq.}}^{id},
\end{equation}
 and

\begin{equation}
\mu ^{(2)}=\frac{kT}{m}\ln \left( \frac{g^{(2)}}{g_{eq.}^{(2)}}\right) +\mu _{B_{eq}}^{exc}.
\end{equation}

After substitution of these expressions in the entropy functional (17), using
the definition of mass density \( \rho _{_{B}}=m\int f^{(1)}d\mathbf{u}_{1} \)
for the Brownian particles and considering that the heat bath density \( \rho _{_{H}} \)
is constant\( , \) we can identify through the Gibbs entropy postulate (16)
the equilibrium entropy,

\begin{equation}
\delta S_{eq}=-\frac{\mu _{B_{eq}}^{tot}}{T}\delta \rho _{_{B}}.
\end{equation}
 This means that eq. (17) with (24) and (25) is consistent with the\ Gibbs entropy
postulate (16) formulated at the pair correlation level.

We now proceed to analyze the evolution in time of Gibbs local entropy \( S(\mathbf{r}_{1},t) \).
To reach this purpose, from eq. (17) we obtain the rate of change of entropy
per unit volume,

\begin{equation}
\frac{\partial }{\partial t}S(\mathbf{r}_{1},t)=-mN\dint \frac{\mu ^{(1)}}{T}\frac{\partial }{\partial t}P^{(1)}d\mathbf{u}_{1}-m^{2}\frac{N}{2}\left( N-1\right) \int \frac{\mu ^{(2)}}{T}\frac{\partial }{\partial t}P^{(2)}d\mathbf{u}_{1}d\mathbf{u}_{2}d\mathbf{r}_{2}.
\end{equation}
 Working out the first term in this expression, after substituting eq. (9) we
have, 
\begin{equation}
-m\dint \frac{\mu ^{(1)}}{T}\frac{\partial }{\partial t}P^{(1)}d\mathbf{u}_{1}=-\frac{\partial }{\partial \mathbf{r}_{i}}\cdot \mathbf{J}_{s}^{(1)}(\mathbf{r}_{1},t)+\sigma ^{(1)}(\mathbf{r}_{1},t).
\end{equation}
 Here\ \ 

\bigskip

\begin{eqnarray}
\mathbf{J}_{s}^{(1)}(\mathbf{r}_{1},t) & = & Nk\int \mathbf{u}_{1}P^{(1)}\left( \frac{m}{kT}\mu ^{(1)}-1\right) d\mathbf{u}_{1}\nonumber \\
 &  & -\frac{N\left( N-1\right) }{2T}\int \left( \frac{\partial \phi _{12}}{\partial r_{_{12}}}\right) \frac{\mathbf{r}_{12}\mathbf{r}_{12}}{r_{12}}\cdot \int _{0}^{1}P^{(2)}d\alpha \frac{\partial \mu ^{(1)}}{\partial \mathbf{u}_{1}}d\mathbf{r}_{12}d\mathbf{u}_{2}d\mathbf{u}_{1}\nonumber \\
 &  & 
\end{eqnarray}
 is the first contribution to the entropy flux, with \( \int _{0}^{1}P^{(2)}d\alpha  \)
\( = \) \( \int _{0}^{1}P^{(2)}\left( \mathbf{r}_{1}-\left[ 1-\alpha \right] ,\right.  \)
\( \left. \mathbf{r}_{1j},\allowbreak \mathbf{u}_{1},\mathbf{r}_{1}+\alpha \mathbf{r}_{1j},\mathbf{u}_{2,}t\right) d\alpha , \)
as results from assuming that \( P^{(2)} \) is a slow varying function of \( \mathbf{r}_{1} \)
and it admits a Taylor's expansion around the distance separation between particles
\( \mathbf{r}_{1j} \) \cite{kirkwood}\( . \) The first contribution to the
entropy production is then given by,

\begin{equation}
\sigma ^{(1)}(\mathbf{r}_{1},t)=-Nk\dint \mathbf{J}_{u_{1}}^{(1)}\cdot \frac{\partial }{\partial \mathbf{u}_{1}}\ln \left( \frac{P^{(1)}}{P_{l.e.}^{(1)}}\right) d\mathbf{u}_{1}.
\end{equation}

We now analyze\ the two-particle term in eq. (27) using eq. (10). After some
algebra we obtain the next identity

\begin{equation}
-m^{2}\int \frac{\mu ^{(2)}}{T}\frac{\partial }{\partial t}P^{(2)}d\mathbf{u}_{1}d\mathbf{u}_{2}d\mathbf{r}_{2}=-\frac{\partial }{\partial \mathbf{r}_{i}}\cdot \mathbf{J}_{s}^{(2)}(\mathbf{r}_{1},t)+\sigma ^{(2)}(\mathbf{r}_{1},t),
\end{equation}
 where

\begin{eqnarray}
\mathbf{J}_{s}^{(2)}(\mathbf{r}_{1},t) & = & -\frac{N\left( N-1\right) m^{2}}{2T}\int P^{(2)}\mathbf{u}_{1}\mu ^{(2)}d\mathbf{u}_{1}d\mathbf{u}_{2}d\mathbf{r}_{2}\nonumber \\
 &  & +\frac{N\left( N-1\right) m^{2}}{2T}\int \left( \mathbf{u}_{1}+\mathbf{u}_{2}\right) \cdot \frac{\mathbf{r}_{12}\mathbf{r}_{12}}{2r_{12}}\frac{\partial }{\partial r_{12}}\mu ^{(2)}\int _{0}^{1}P^{(2)}d\alpha d\mathbf{u}_{1}d\mathbf{u}_{2}d\mathbf{r}_{12}\nonumber \\
 &  & -\frac{N\left( N-1\right) }{2T}\int \sum _{i,j=1}^{2}\frac{\partial \phi _{ij}}{\partial r_{ij}}\frac{\mathbf{r}_{ij}\mathbf{r}_{ij}}{r_{ij}}\cdot \int _{0}^{1}P^{(2)}d\alpha \frac{\partial \mu ^{(2)}}{\partial \mathbf{u}_{1}}d\mathbf{u}_{1}d\mathbf{u}_{2}d\mathbf{r}_{ij},\nonumber \\
 &  & 
\end{eqnarray}
 is the two-particle entropy flux and

\begin{eqnarray}
\sigma ^{(2)}(\mathbf{r}_{1},t) & = & -\frac{N\left( N-1\right) }{2}k\int \mathbf{J}_{u_{1}}^{(2)}\cdot \frac{\partial }{\partial \mathbf{u}_{1}}\ln \left( \frac{g^{(2)}}{g_{eq}^{(2)}}\right) d\mathbf{u}_{1}d\mathbf{u}_{2}d\mathbf{r}_{2}\nonumber \\
 &  & -\frac{N\left( N-1\right) }{2}k\int \mathbf{J}_{u_{2}}^{(2)}\cdot \frac{\partial }{\partial \mathbf{u}_{2}}\ln \left( \frac{g^{(2)}}{g_{eq}^{(2)}}\right) d\mathbf{u}_{1}d\mathbf{u}_{2}d\mathbf{r}_{2,}
\end{eqnarray}
 is the corresponding two-particle entropy production.

A previous analysis on simultaneous motion of \( N \)-Brownian particles leads
to the next global entropy production \cite{rubi},

\begin{equation}
\sigma (t)=\int \sigma ^{(N)}d\mathbf{x}^{N}=\sum _{i=1}^{N}\int \mathbf{J}_{u_{i}}^{(N)}\cdot k\frac{\partial }{\partial \mathbf{u}_{i}}\ln \left( \frac{P^{(N)}}{P_{eq}^{(N)}}\right) d\mathbf{x}^{N}.
\end{equation}
 Inserting in this expression the distribution functions \( \ln P^{(N)} \)
and \( \ln P_{eq}^{(N)}, \) given through eqs. (14) and (15) we obtain an alternative
expression for the global entropy production, i.e.,

\[
\int \sigma ^{\left( N\right) }d\mathbf{x}^{N}=\]

\begin{eqnarray}
 & = & Nk\int \mathbf{J}_{u_{1}}^{(1)}\cdot \frac{\partial }{\partial \mathbf{u}_{1}}\ln \left( \frac{P^{(1)}}{P_{eq}^{(1)}}\right) d\mathbf{u}_{1}d\mathbf{r}_{1}+\nonumber \\
 &  & \frac{N\left( N-1\right) }{2}k\sum _{j=1}^{2}\int \mathbf{J}_{u_{i}}^{\left( 2\right) }\cdot \frac{\partial }{\partial \mathbf{u}_{i}}\ln \left( \frac{g^{(2)}}{g_{eq.}^{(2)}}\right) d\mathbf{u}_{2}d\mathbf{r}_{12}d\mathbf{u}_{1}+...
\end{eqnarray}
 Where we have taken into account the definition of the reduced fluxes, eq.
(8).

We notice that the local entropy production \( \sigma \left( \mathbf{r}_{1},t\right) =\sigma ^{\left( 1\right) }\left( \mathbf{r}_{1},t\right) + \)\ \( \sigma ^{\left( 2\right) }\left( \mathbf{r}_{1},t\right)  \)
can be identified in the integrand of eq. (35). This means that consistently
we can use the integrand of the eq. (34) as the local entropy production, and
afterwards the hypothesis \ that the fluxes \( \mathbf{J}_{u_{i}}^{(N)} \)
are coupled to the thermodynamic forces \( \frac{\partial }{\partial \mathbf{u}_{i}}\ln \left( \frac{P^{(N)}}{P_{eq}^{(N)}}\right) , \)
giving rise to the next phenomenological equations,

\begin{equation}
\mathbf{J}_{u_{i}}^{(N)}=-k\sum \Sb j\neq ij=1\endSb ^{N}L_{_{u_{i}u_{j}}}\frac{\partial }{\partial \mathbf{u}_{i}}\ln \left( \frac{P^{(N)}}{P_{eq}^{(N)}}\right) .
\end{equation}

Defining the friction coefficients as

\begin{equation}
\beta _{_{ij}}=\frac{mL_{u_{i}u_{j}}}{P^{(N)}T_{eq}},
\end{equation}
 we can give an alternative and more useful expression for the fluxes \( \mathbf{J}_{u_{i}}^{(N)} \),
namely,

\begin{equation}
\mathbf{J}_{u_{i}}^{(N)}=-\sum \Sb j\neq ij=1\endSb ^{N}\beta _{ij}\left( P^{(N)}\mathbf{u}_{j}+\frac{kT_{eq}}{m}\frac{\partial P^{(N)}}{\partial \mathbf{u}_{j}}\right) .
\end{equation}
 Particularly, for the one and two-particle description the corresponding fluxes
are,

\begin{equation}
\mathbf{J}_{u_{1}}^{(1)}=\beta _{11}\left( P^{(1)}\mathbf{u}_{1}+\frac{kT_{eq}}{m}\frac{\partial P^{(1)}}{\partial \mathbf{u}_{1}}\right) ,
\end{equation}
 and

\begin{equation}
\mathbf{J}_{u_{i}}^{(2)}=\sum _{i,j=1}^{2}\beta _{_{i}{}_{j}}\left( P^{(2)}\mathbf{u}_{i}+\frac{kT_{eq}}{m}\frac{\partial P^{(2)}}{\partial \mathbf{u}_{i}}\right) .
\end{equation}

Inserting now these expressions into the continuity equations (9) and (10) one
obtains the desired Fokker-Planck equations,

\[
\frac{\partial }{\partial t}P^{(1)}+\mathbf{u}_{1}\cdot \frac{\partial }{\partial \mathbf{r}_{1}}P^{(1)}-m^{-1}\int \frac{\partial \phi _{12}}{\partial \mathbf{r}_{1}}\cdot \frac{\partial P^{(2)}}{\partial \mathbf{u}_{1}}d\mathbf{u}_{2}d\mathbf{r}_{2}\]

\begin{equation}
=\frac{\partial }{\partial \mathbf{u}_{1}}\cdot \beta _{11}\left( P^{(1)}\mathbf{u}_{1}+\frac{kT_{eq}}{m}\frac{\partial P^{(1)}}{\partial \mathbf{u}_{1}}\right) 
\end{equation}
 and

\[
\frac{\partial }{\partial t}P^{(2)}+\sum _{i=1}^{2}\mathbf{u}_{i}\cdot \frac{\partial P^{(2)}}{\partial \mathbf{r}_{i}}-m^{-1}\sum _{i,j=1}^{2}\frac{\partial \phi _{ij}}{\partial \mathbf{r}_{i}}\cdot \frac{\partial P^{(2)}}{\partial \mathbf{u}_{i}}\]

\begin{equation}
=\sum _{i,j=1}^{2}\frac{\partial }{\partial \mathbf{u}_{i}}\cdot \beta _{_{i}{}_{j}}\left( P^{(2)}\mathbf{u}_{i}+\frac{kT_{eq}}{m}\frac{\partial P^{(2)}}{\partial \mathbf{u}_{i}}\right) .
\end{equation}

The analysis of these equations, deserves some comments. The third term on the
left hand side of eq. (41) represents the local equilibrium mean effective interaction
potential applied at the point \( \mathbf{r}_{1} \), where we have included
the instantaneous interaction of a Brownian particle placed at \( \mathbf{r}_{1}, \)
with the solvent. If we do not take into account this mean field term, coming
from the interactions of a cloud of \ Brownian particles around \( \mathbf{r}_{1}, \)
including the local structure of the solvent, we recover the usual one-particle
Fokker-Planck equation \cite{dhont}.

With respect to eq. (42), the effective interactions \( -m^{-1}\sum _{i,j=1}^{2}\frac{\partial \phi _{ij}}{\partial \mathbf{r}_{i}}\cdot \frac{\partial P^{(2)}}{\partial \mathbf{u}_{i}} \)
are applied in a non-local manner over the two tracer particles placed at \( \mathbf{r}_{1} \)
and \( \mathbf{r}_{2}, \) respectively. The friction tensor is taken by pairs
now with respect to the two selected particles. This Fokker-Planck equation
coincides with the one obtained first by Mazo \cite{mazo} and more recently
by Piasecki et al \cite{piasecki}. This last one was derived for a system of
two-Brownian hard spheres immersed in a hard-sphere solvent.

When the velocities of the Brownian particles thermalize the system reaches
the state of local equilibrium. For this case, the dynamical description of
the Brownian particles would be represented by the evolution equations for the
one- and two- particle local equilibrium distribution functions. Hence, the
right hand sides of the continuity equations ( 9) and (10) should be replaced
by \( -\frac{\partial \mathbf{J}_{r_{1}}^{\left( 1\right) }}{\partial \mathbf{r}_{1}} \),
and \( -\sum _{i=1}^{2}\frac{\partial \mathbf{J}_{r_{i}}^{\left( 2\right) }}{\partial \mathbf{r}_{i}} \),
respectively. Following a similar analysis as the one performed in this section,
we would obtain the expressions for the fluxes \( \mathbf{J}_{r_{1}}^{\left( 1\right) } \)
and \( \mathbf{J}_{r_{i}}^{\left( 2\right) } \). The final result would correspond
to a couple of equations of Smoluchowski type. Such approach has allowed the
calculation of the long time self-diffusion coefficient of hard sphere suspensions
and, the corresponding comparison with Brownian dynamics simulations \cite{leegwater}.

The one- and two-particles Fokker-Planck equations describe the dynamics of
the system at mesoscopic level, under the assumption of pair interactions. These
expressions can be used to analyze the hydrodynamic regime of the Brownian gas.
This will be the goal of the next section.

\section{\protect\smallskip \protect\smallskip Hydrodynamic equations}

\qquad Our purpose in this section is to derive the complete set of hydrodynamic
equations describing macroscopically the dynamics of the gas of Brownian particles.
For simplicity in our notation, we will omit the dependences of the different
quantities in \( (\mathbf{r}_{1},t). \) The conservation law for the mass of
the particles follows from the definition of the density

\begin{equation}
\rho _{_{B}}=m\int P^{(1)}d\mathbf{u}_{1}.
\end{equation}
 ~

\noindent We take the temporal derivative of this expresion and use the continuity
equation (9). After integrating by parts the resulting equation and using the
fact that the current \( \mathbf{J}_{u_{1}}^{(1)} \) decays very rapidly when
increasing the velocity, one obtains the conservation law

\begin{equation}
\frac{\partial \rho _{_{B}}}{\partial t}=-\frac{\partial }{\partial \mathbf{r}_{1}}\cdot \rho _{_{B}}\mathbf{v}_{B}.
\end{equation}

We can proceed in a similar way\cite{perezmadrid} with the definition of the
momentum density of the Brownian particles

\begin{equation}
\rho _{_{B}}\mathbf{v}_{B}=m\int \mathbf{u}_{1}P^{(1)}d\mathbf{u}_{1}.
\end{equation}
 By taking the time derivative of this equation and using the one-particle Fokker-Planck
equation (41), we obtain a preliminar balance equation for the momentum density,

\[
\frac{\partial \rho _{_{B}}\mathbf{v}_{B}}{\partial t}+\frac{\partial }{\partial \mathbf{r}_{1}}\cdot \left( \overleftrightarrow {\mathcal{P}}_{B}^{k}+\rho _{_{B}}\mathbf{v}_{B}\mathbf{v}_{B}\right) =\]

\begin{equation}
-\int P^{(2)}\frac{\partial \phi _{12}}{\partial \mathbf{r}_{1}}d\mathbf{r}_{2}d\mathbf{u}_{2}d\mathbf{u}_{1}-\beta _{11}\rho _{_{B}}\mathbf{v}_{B}.
\end{equation}
 Here

\begin{equation}
\overleftrightarrow {\mathcal{P}}_{B}^{k}=m\int P^{(1)}\left( \mathbf{u}_{1}-\mathbf{v}_{B}\right) \left( \mathbf{u}_{1}-\mathbf{v}_{B}\right) d\mathbf{u}_{1},
\end{equation}
 is the kinetic part of the pressure tensor for the Brownian particles, with
\( \beta _{11} \) the friction coeficient, such that \( \beta _{11}\rho _{_{B}}\mathbf{v}_{B} \)
takes into account the momentum exchange between the Brownian particles and
the bath.

Now, we admit that the distribution function \( P^{(2)} \) varies slowly with
the space coordinates as it should be for a concentrated suspension. Expanding
the probability density \( P^{(2)} \) around the distance \( \mathbf{r}_{12}=\mathbf{r}_{2}-\mathbf{r}_{1} \)\textbf{\ {}}between
particles\textbf{\ {}} \cite{ferziger},\textbf{\ {}}we may approximate

\[
\int P^{(2)}\frac{\partial \phi _{12}}{\partial \mathbf{r}_{1}}d\mathbf{r}_{2}\approx \]

\begin{equation}
\approx -\frac{\partial }{\partial \mathbf{r}_{1}}\cdot \frac{m}{2}\int \mathbf{r}_{12}\mathbf{r}_{12}\frac{\phi _{_{12}}^{^{\prime }}\left( r\right) }{r_{12}}\int _{0}^{1}P^{(2)}\left( \mathbf{r}_{1}-\left[ 1-\alpha \right] \mathbf{r}_{12},\mathbf{r}_{1}+\alpha \mathbf{r}_{12},\mathbf{u}_{1,}\mathbf{u}_{2,}t\right) d\alpha d\mathbf{r}_{12},
\end{equation}
 with \( \phi _{12}^{^{^{\prime }}}\left( r_{_{12}}\right) =\frac{\partial \phi _{12}}{\partial r_{12}} \)
\( . \) This relation enables us to identify the potential component of the
pressure tensor, namely,

\begin{eqnarray}
\overleftrightarrow {\mathcal{P}}_{B}^{^{\phi }} & = & -\frac{m}{2}\int \mathbf{r}_{12}\mathbf{r}_{12}\frac{\phi _{12}^{^{\prime }}\left( r_{12}\right) }{r_{12}}\times \nonumber \\
 &  & \left( \int _{0}^{1}P^{(2)}\left( \mathbf{r}_{1}-\left[ 1-\alpha \right] \mathbf{r}_{12},\mathbf{r}_{1}+\alpha \mathbf{r}_{12},\mathbf{u}_{1},\mathbf{u}_{2},t\right) d\alpha \right) d\mathbf{r}_{12}d\mathbf{u}_{1}d\mathbf{u}_{2}.\nonumber \\
 &  & 
\end{eqnarray}

By means of the above identity, the momentum balance (46) can be rewritten in
the usual manner \cite{kirkwood,kreuzer},

\begin{equation}
\rho _{_{B}}\frac{d\mathbf{v}_{B}}{dt}=-\frac{\partial }{\partial \mathbf{r}_{1}}\cdot \overleftrightarrow {\mathcal{P}_{B}}-\beta _{11}\rho _{_{B}}\mathbf{v}_{B},
\end{equation}
 with

\begin{equation}
\overleftrightarrow {\mathcal{P}}_{B}=\overleftrightarrow {\mathcal{P}}_{B}^{k}+\overleftrightarrow {\mathcal{P}}_{B}^{^{\phi }},
\end{equation}
 the total pressure tensor of the Brownian particles and \( \frac{d}{dt}=\frac{\partial }{\partial t}+\mathbf{v}_{B}\cdot \frac{\partial }{\partial \mathbf{r}_{1}} \)
the hydrodynamic derivative. This expression for the pressure tensor is consistent
with the one obtained by Felderhof \cite{felderhof1} (eqs. 3.11 and 3.14 of
such reference), who used the N-particle Fokker-Planck equation for its derivation.

In order to complete our scheme, we need to obtain the balance equation for
the density of internal energy of the ``gas'' of Brownian particles \( \rho _{_{B}}u_{_{B}} \).
This quantity splits up into kinetic and potential contributions

\begin{equation}
\rho _{_{B}}u_{_{B}}=\rho _{_{B}}u_{_{B}}^{^{k}}+\rho _{_{B}}u_{_{B}}^{^{\phi }},
\end{equation}
 where

\begin{equation}
\rho _{_{B}}u_{_{B}}^{^{k}}=\frac{m}{2}\int P^{(1)}\left( \mathbf{u}_{1}-\mathbf{v}_{B}\right) ^{2}d\mathbf{u}_{1}
\end{equation}
 and

\begin{equation}
\rho _{_{B}}u_{_{B}}^{^{\phi }}=\frac{m}{2}\int P^{(2)}\phi _{12}d\mathbf{u}_{1}d\mathbf{u}_{2}d\mathbf{r}_{2},
\end{equation}
 are the densities of kinetic and potential energies, respectively. Taking the
temporal derivative of eq. (53) and using the Fokker-Planck eq. (41), one obtains,

\[
\frac{\partial }{\partial t}\rho _{_{B}}u_{_{B}}^{^{k}}=-\frac{\partial }{\partial \mathbf{r}_{1}}\cdot \frac{m}{2}\int \left( \mathbf{u}_{1}-\mathbf{v}_{B}\right) ^{2}\mathbf{u}_{1}P^{(1)}d\mathbf{u}_{1}+\frac{m}{2}\int P^{(1)}\mathbf{u}_{1}\cdot \frac{\partial \left( \mathbf{u}_{1}-\mathbf{v}_{B}\right) ^{2}}{\partial \mathbf{r}_{1}}d\mathbf{u}_{1}\]

\[
+\frac{m}{2}\int \left( \mathbf{u}_{1}-\mathbf{v}_{B}\right) ^{2}\frac{\partial }{\partial \mathbf{u}_{1}}\cdot \beta _{_{11}{}}\left[ P^{(1)}\mathbf{u}_{1}+\frac{kT_{eq}}{m}\frac{\partial P^{(1)}}{\partial \mathbf{u}_{1}}\right] d\mathbf{u}_{1}\]

\[
=-\frac{\partial }{\partial \mathbf{r}_{1}}\cdot \left[ \frac{m}{2}\int \left( \mathbf{u}_{1}-\mathbf{v}_{B}\right) ^{2}\left( \mathbf{u}_{1}-\mathbf{v}_{B}\right) P^{(1)}d\mathbf{u}_{1}+\frac{m}{2}\int \left( \mathbf{u}_{1}-\mathbf{v}_{B}\right) ^{2}P^{(1)}d\mathbf{u}_{1}\mathbf{v}_{B}\right] \]

\[
+\frac{m}{2}\int P^{(1)}\mathbf{u}_{1}\cdot \frac{\partial }{\partial \mathbf{r}_{1}}\left( \mathbf{u}_{1}-\mathbf{v}_{B}\right) ^{2}d\mathbf{u}_{1}\]

\[
-m\left( \frac{T-T_{eq}}{T}\right) \beta _{_{11}{}}\int P^{(1)}\left( \mathbf{u}_{1}-\mathbf{v}_{B}\right) ^{2}d\mathbf{u}_{1}.\]
 Defining now

\begin{equation}
\mathbf{J}_{q}^{k}=\frac{m}{2}\int \left( \mathbf{u}_{1}-\mathbf{v}_{B}\right) ^{2}\left( \mathbf{u}_{1}-\mathbf{v}_{B}\right) P^{(1)}d\mathbf{u}_{1}
\end{equation}
 the kinetic part of the heat flux, using the definition of the kinetic energy
density \ (53) and the next identity,

\[
\frac{m}{2}\int P^{(1)}\mathbf{u}_{1}\cdot \frac{\partial }{\partial \mathbf{r}_{1}}\left( \mathbf{u}_{1}-\mathbf{v}_{B}\right) ^{2}d\mathbf{u}_{1}=\]

\begin{eqnarray*}
 & = & m\int P^{(1)}\mathbf{u}_{1}\left( \mathbf{u}_{1}-\mathbf{v}_{B}\right) :\frac{\partial \left( \mathbf{u}_{1}-\mathbf{v}_{B}\right) }{\partial \mathbf{r}_{1}}d\mathbf{u}_{1}\\
 & = & -\left[ m\int P^{(1)}\left( \mathbf{u}_{1}-\mathbf{v}_{B}\right) \left( \mathbf{u}_{1}-\mathbf{v}_{B}\right) d\mathbf{u}_{1}\right] :\frac{\partial \mathbf{v}_{B}}{\partial \mathbf{r}_{1}},
\end{eqnarray*}
 we obtain the balance equation for the kinetic part,

\[
\frac{\partial }{\partial t}\rho _{_{B}}u_{_{B}}^{^{k}}=-\overleftrightarrow {\mathcal{P}}_{B}^{k}:\frac{\partial \mathbf{v}_{B}}{\partial \mathbf{r}_{1}}-\frac{\partial }{\partial \mathbf{r}_{1}}\cdot \left( \mathbf{J}_{q}^{k}-\rho _{_{B}}u_{_{B}}^{^{k}}\mathbf{v}_{B}\right) \]

\begin{equation}
-m\left( \frac{T-T_{eq}}{T}\right) \beta _{_{11}{}}\int P^{(1)}\left( \mathbf{u}_{1}-\mathbf{v}_{B}\right) ^{2}d\mathbf{u}_{1}.
\end{equation}

The derivation of the above balance equation deserves some comments. First of
all we have assumed that the friction coefficient is independent of the velocity.
This means that we have not taken into account the non-linear friction case.
The velocity dependence of the friction coefficient \( \beta _{_{11}}\left( \mathbf{u}_{1}\right)  \)
concerns with the case in which the Brownian particles are able to take up external
energy, which can be stored in an internal energy depot \cite{erdmann}. This
phenomenon is the so-called active motion and is of interest in the dynamical
analysis of driven physico-chemical \cite{mikhailov} and biological systems
\cite{schienbein}. If we want to take into account this fact, the above balance
equation (56) would include the additional term \( -m\int P^{(1)}\beta _{_{1}{}_{1}}\left( \mathbf{u}_{1}-\mathbf{v}_{B}\right) \cdot \mathbf{v}_{B}d\mathbf{u}_{1}. \)
Moreover, the temperature \( T \) corresponds to a non-equilibrium distribution
of the Brownian particles, which eventually thermalizes to the usual Maxwell
velocity distribution function \( P_{eq}^{(1)} \)with the equilibrium temperature
\( T_{eq}. \) In this manner, the last term of \ eq. (56) accounts for the
kinetic energy interchange between the Brownian particles and the heat bath.

Using the kinetic energy density definition (53) together with the energy equipartition
theorem, namely,

\begin{equation}
\rho _{_{B}}u_{_{B_{eq}}}^{^{k}}=\frac{3}{2}\frac{\rho _{_{B}}}{m}kT_{eq},
\end{equation}
 the last term in the equation (56) can be rewritten as follows,

\begin{equation}
-m\left( \frac{T-T_{eq}}{T}\right) \beta _{_{11}{}}\int P^{(1)}\left( \mathbf{u}_{1}-\mathbf{v}_{B}\right) ^{2}d\mathbf{u}_{1}=-2\rho _{_{B}}\beta _{_{11}{}}\left( u_{_{B}}^{^{k}}-u_{_{B_{eq}}}^{^{k}}\right) .
\end{equation}

With this expression, we recover the usual kinetic energy balance equation for
a dilute suspension of Brownian particles \cite{degroot},

\begin{equation}
\frac{\partial }{\partial t}\rho _{_{B}}u_{_{B}}^{^{k}}=-\overleftrightarrow {\mathcal{P}}_{B}^{k}:\frac{\partial \mathbf{v}_{B}}{\partial \mathbf{r}_{1}}-\frac{\partial }{\partial \mathbf{r}_{1}}\left( \mathbf{J}_{q}^{k}-\rho _{_{B}}u_{_{B}}^{^{k}}\mathbf{v}_{B}\right) -2\beta _{_{11}{}}\rho _{_{B}}\left( u_{_{B}}^{^{k}}-u_{_{B_{eq}}}^{^{k}}\right) 
\end{equation}

The potential energy balance equation, follows from its definition eq.(54).
By using the two-particle Fokker-Planck eq.(42) and Newton's third law \( \frac{\partial \phi _{12}}{\partial \mathbf{r}_{1}}=-\frac{\partial \phi _{12}}{\partial \mathbf{r}_{2}} \),
we obtain

\begin{equation}
\frac{\partial }{\partial t}\rho _{_{B}}u_{_{B}}^{^{\phi }}=-\frac{\partial }{\partial \mathbf{r}_{1}}\cdot \left( \mathbf{J}_{q\phi }^{(1)}+\rho _{_{B}}u_{_{B}}^{^{\phi }}\mathbf{v}_{B}\right) +\frac{m}{2}\int P^{(2)}\left( \mathbf{u}_{1}-\mathbf{u}_{2}\right) \cdot \frac{\partial \phi _{12}}{\partial \mathbf{r}_{1}}d\mathbf{u}_{1}d\mathbf{u}_{2}d\mathbf{r}_{2},
\end{equation}
 in which

\begin{equation}
\mathbf{J}_{q\phi }^{(1)}\left( \mathbf{r}_{1},t\right) =m\int \left( \mathbf{u}_{1}-\mathbf{v}_{B}\right) \frac{\phi _{12}}{2}P^{(2)}d\mathbf{u}_{1}d\mathbf{u}_{2}d\mathbf{r}_{2},
\end{equation}
 is the current of potential energy density transport and \( \frac{m}{2}\int P^{(2)}\left( \mathbf{u}_{1}-\mathbf{u}_{2}\right) \cdot \frac{\partial \phi _{12}}{\partial \mathbf{r}_{1}}d\mathbf{u}_{1}d\mathbf{u}_{2}d\mathbf{r}_{2} \)
is the production term. After applying the approximate expression (48) one has,

\[
\frac{\partial }{\partial t}\rho _{B}u_{_{B}}^{^{\phi }}=-\frac{\partial }{\partial \mathbf{r}_{1}}\cdot \left( \mathbf{J}_{q\phi }^{(1)}+\mathbf{J}_{q\phi }^{(2)}+\rho _{B}u_{_{B}}^{^{\phi }}\mathbf{v}_{B}\right) \]

\begin{equation}
+\int P^{(2)}\frac{\partial \phi _{12}}{\partial \mathbf{r}_{1}}\cdot \mathbf{u}_{1}d\mathbf{u}_{1}d\mathbf{u}_{2}d\mathbf{r}_{2}-\overleftrightarrow {\mathcal{P}}_{B}^{^{\phi }}:\frac{\partial \mathbf{v}_{B}}{\partial \mathbf{r}_{1}},
\end{equation}
 where

\[
\mathbf{J}_{q\phi }^{(2)}\left( \mathbf{r}_{1},t\right) =-\frac{1}{4}\int \mathbf{r}_{12}\mathbf{r}_{12}\frac{\phi _{12}^{\prime }\left( r_{12}\right) }{r_{12}}\left( \mathbf{u}_{1}+\mathbf{u}_{2}-2\mathbf{v}_{B}\right) \times \]

\begin{equation}
\times \int _{0}^{1}mP^{(2)}\left( \mathbf{r}_{1}-\left[ 1-\alpha \right] \mathbf{r}_{12},\mathbf{u}_{2},\mathbf{r}_{1}+\alpha \mathbf{r}_{12},\mathbf{u}_{1};t\right) d\alpha d\mathbf{u}_{1}d\mathbf{u}_{2}d\mathbf{r}_{12}
\end{equation}
 is the potential contribution to the heat transport.

Regarding the term \( \int P^{(2)}\frac{\partial \phi _{12}}{\partial \mathbf{r}_{1}}\cdot \mathbf{u}_{1}d\mathbf{u}_{1}d\mathbf{u}_{2}d\mathbf{r}_{2}, \)
when we neglect the velocity dependence of the pair dynamic correlation function,
i.e., \( P^{(2)}(\mathbf{x}_{1,}\mathbf{x}_{2},t)=P^{(1)}\left( \mathbf{x}_{1},t\right) P^{(1)}\left( \mathbf{x}_{2},t\right) g^{\left( 2\right) }\left( \mathbf{r}_{1,}\mathbf{r}_{2,}t\right)  \),
we have,

\[
\int P^{(2)}\frac{\partial \phi _{12}}{\partial \mathbf{r}_{1}}\cdot \mathbf{u}_{1}d\mathbf{u}_{1}d\mathbf{u}_{2}d\mathbf{r}_{2}=\]

\[
=\frac{\rho _{_{B}}\left( \mathbf{r}_{1},t\right) }{m}\mathbf{v}_{B}\cdot \int \frac{\rho _{_{B}}\left( \mathbf{r}_{2},t\right) }{m}g^{\left( 2\right) }\left( \mathbf{r}_{1,}\mathbf{r}_{2,}t\right) \frac{\partial \phi _{12}}{\partial \mathbf{r}_{1}}d\mathbf{r}_{2}\]

\begin{equation}
=-\frac{\rho _{_{B}}\left( \mathbf{r}_{1},t\right) }{m}\mathbf{v}_{B}\cdot \mathbf{F}\left( \mathbf{r}_{1},t\right) .
\end{equation}

Taking into account that the potential contributions to the pressure tensor
(eq. 49) and the second part for the heat flux (eq. \ 63) estimate the short
range interactions by means of the approximation described in eq. (48), the
force \( \mathbf{F}\left( \mathbf{r}_{1},t\right) =-\int \frac{\rho _{_{B}}\left( \mathbf{r}_{2},t\right) }{m}g^{\left( 2\right) }\left( \mathbf{r}_{1,}\mathbf{r}_{2,}t\right) \frac{\partial \phi _{12}}{\partial \mathbf{r}_{1}}d\mathbf{r}_{2} \)
accounts for the long range interactions. With these considerations, the potential
energy balance takes the conventional macroscopic form \cite{degroot},

\begin{equation}
\frac{\partial }{\partial t}\rho _{_{B}}u_{_{_{B}}}^{^{\phi }}=-\frac{\partial }{\partial \mathbf{r}_{1}}\cdot \left( \mathbf{J}_{q\phi }^{(1)}+\mathbf{J}_{q\phi }^{(2)}+\rho _{_{B}}u_{_{B}}^{^{\phi }}\mathbf{v}_{B}\right) -\overleftrightarrow {\mathcal{P}}_{B}^{^{\phi }}:\frac{\partial \mathbf{v}_{B}}{\partial \mathbf{r}_{1}}-\frac{\rho _{_{B}}}{m}\mathbf{v}_{B}\cdot \mathbf{F}.
\end{equation}

The balance of internal energy follows just by adding eqs. (59) and (65), i.e,

\[
\frac{\partial }{\partial t}\rho _{_{B}}u_{_{B}}+\frac{\partial }{\partial \mathbf{r}_{1}}\cdot \left( \mathbf{J}_{q}^{B}+\rho _{_{B}}u_{_{B}}\mathbf{v}_{B}\right) =-\frac{\rho _{_{B}}}{m}\mathbf{v}_{B}\cdot \mathbf{F}+\overleftrightarrow {\mathcal{P}}_{B}:\frac{\partial \mathbf{v}_{B}}{\partial \mathbf{r}_{1}}\]

\begin{equation}
-2\rho _{B}\beta _{_{11}{}}\left( u_{_{B}}^{^{k}}-u_{_{B_{eq}}}^{^{k}}\right) ,
\end{equation}
 where

\begin{equation}
\mathbf{J}_{q}^{B}=\mathbf{J}_{qk}+\mathbf{J}_{q\phi }^{(1)}+\mathbf{J}_{q\phi }^{(2)}
\end{equation}
 is the total heat flux, due to the energy interchange of Brownian particles
with the solvent. The term \( -\frac{\rho _{B}}{m}\mathbf{v}_{B}\cdot \mathbf{F} \)
accounts for the rate at which work is done by a Brownian particle over the
solvent due to the static pair effective long range interactions. This term
\ would be useful to account for the active motion mentioned above \cite{erdmann}.

In this manner, with the help of two-particle Fokker-Planck eqs. (41) and (42),
we have obtained the corresponding balance equations for the suspended particles.
The expressions for the pressure tensor (eqs. (47) and (49)), and the heat flux
(eqs. (55),(61) and (63)) are of particular interest. Using a suitable interaction
potential and a pair correlation function for the interacting Brownian particles,
and comparing with Fourier and Newton laws, approximate relationships for the
heat conductivity and viscosity can be obtained. Moreover, with the help of
the momentum\ balance eq. (50), we can analyze the diffusion\ regime. This point
will be the treated in the next section.

\section{Diffusion regime}

\qquad As follows from the balance equation for the momentum of the Brownian
particles, the mobility \( b_{11}=\beta _{11}^{-1} \), introduces a characteristic
time scale defining the inertial regime in the dynamics of the Brownian particles.
For times such that \( t>> \) \( b_{11} \), the particle enters the diffusive
regime in which the inertial term on the left hand side of eq.(50) is negligibly
small, hence this equation yields,

\begin{equation}
-\frac{\partial }{\partial \mathbf{r}_{1}}\cdot \overleftrightarrow {\mathcal{P}}_{B}=\beta _{11}\rho _{_{B}}\mathbf{v}_{B}.
\end{equation}

For the particular case where no external flow is imposed to the system, the
pressure tensor \( \overleftrightarrow {\mathcal{P}}_{B} \) reduces to the
osmotic pressure \( \overleftrightarrow {\mathcal{P}}_{B}=p_{_{B}}\overleftrightarrow {\mathcal{U}} \),
with \( \overleftrightarrow {\mathcal{U}} \) being the unit tensor and 
\begin{equation}
p_{_{B}}=\frac{\rho _{_{B}}}{m}kT-\frac{2\pi }{3}\left( \frac{\rho _{_{B}}}{m}\right) ^{2}\int \frac{\partial \phi _{12}}{\partial r}g_{eq}^{(2)}(r_{12})r_{_{12}}^{^{3}}dr_{12},
\end{equation}
 the pressure that the set of Brownian particles exerts on the host fluid, recovering
the usual virial equation of state. This result agrees with the one obtained
by Brady using purely hydrodynamic arguments \cite{brady}.

\noindent Using the definition of the mass current \( \mathbf{J}=\rho _{_{B}}\mathbf{v}_{B} \)
and the expression of the pressure (69), we obtain from eq. (68) the Fick's
law

\begin{equation}
\mathbf{J}=-D_{c}\frac{\partial \rho _{_{B}}}{\partial \mathbf{r}},
\end{equation}
 from which we can identify the collective diffusion coefficient. 
\begin{equation}
D_{c}=\frac{kT}{m}\left[ 1-\frac{4}{3}\pi \frac{\rho _{_{B}}}{kTm}\int _{0}^{\infty }\frac{\partial \phi _{12}}{\partial r}g_{eq}^{(2)}r_{_{12}}^{^{3}}dr_{12}\right] b_{11}.
\end{equation}

For long times, this quantity accounts for the relaxation of the spatial distribution
of Brownian particles. In the redistribution of particles, we have included
the influence of the solvent through the additional contribution to the potential
interaction per particle\( . \) For the particular case in which the effective
interactions related to \( \phi _{12} \) decay faster than \( r_{_{12}}^{^{-3}} \),
where \( r_{12} \) is the relative separation of two Brownian particles, the
diffusion coefficient becomes 
\begin{equation}
D_{c}=\frac{kT}{m}\left[ 1-4\pi \frac{\rho _{B}}{m}\int\limits _{0}^{\infty }\left\{ g_{eq}^{(2)}(r_{12})-1\right\} r_{_{12}}^{^{2}}dr_{12}\right] b_{11},
\end{equation}
 which agrees with the corresponding expression given by Russel \cite{russel2}.

Alternatively, by using the virial equation of state for the Brownian particles
(69), we obtain the usual thermodynamic relationship \cite{dhont}, \cite{russel},
between the macroscopic collective diffusion tensor and the osmotic compressibility,

\noindent 
\begin{equation}
D_{c}=b_{11}\left( \frac{\partial p_{B}}{\partial \rho _{B}}\right) _{T},
\end{equation}
 The ensuing expression relating collective diffusion, mobility and compressibility,
deserves some additional comments. First of all, it relates the mass transfer
ratio with the isothermal compressibility as the driven force and the hydrodynamic
mobility as the response function. The mobility appears in our scheme as a phenomenological
coefficient,which may in general depend on time\cite{bonet}. The explicit expression
of this coefficient must be borrowed from hydrodynamics.

 At the two-particle level of our description, the appropriate approach
is the one given by Batchelor \cite{batchelor}. Moreover, the pressure tensor
(51) arising from the momentum balance accounts for the pressure exerted over
the solvent, thus the pressure difference between two separate points drives
the collective motion of Brownian particles from different sides of the solvent.
This fact has been used to interpret light scattering experiments in micellar
emulsions\ \cite{agterof},\cite{cazabat} and lyophilic silica particles in
nonpolar solvents\ \cite{vanhelden} where eq. (71) and the virial form for
the pressure were used as a working hypothesis. It\ has served to test the application
of liquid theories to \ suspensions and to analyze its validity for arbitrary
concentrations and for pair continuous interaction potentials \cite{russel2}\cite{russel}.

The above relation must be complemented with the functional dependence of \( b_{11}\left( \Phi \right)  \)
and \( p_{B}\left( \Phi \right) , \) on \( \Phi =\frac{4}{3}\pi a^{3}\frac{\rho _{B}}{m} \)
the volume fraction of the Brownian particles. The expression for the mobility
\( b_{11}\left( \Phi \right)  \) can again be obtained from hydrodynamics.
For a hard sphere model with stick boundary conditions in the dilute regime,
the result is \( b_{11}\left( \Phi \right) =\left( 6\pi \eta a\right) ^{-1}\left( 1-6.55\Phi \right)  \)
\cite{batchelor}, with \( \eta  \) the viscosity of the fluid phase and \( a \)
the radius of the Brownian particle. An analysis that makes use of a combination
of stick and slip boundary conditions\cite{felderhof2} helps to fit experimental
data\cite{cazabat}. Moreover, for strongly charged spheres at low salinity
based on an effective macroion fluid theoretical model, the mobility is well\
represented by the parametric form \( b_{11}\left( \Phi \right) =\left( 6\pi \eta a\right) ^{-1}\left( 1-p\Phi ^{\alpha }\right)  \)
\cite{watzlawek}. Using a model of effective hard \ spheres with \( \Phi  \)
-dependent diameter, the values \( p\simeq 1.8 \) and \( \alpha =\frac{1}{3} \),
can be explained.

The effect of interactions in the Brownian motion, was discussed in a phenomenological
manner by Van den Broeck et al \cite{broeck} through the generalized Einstein
expression (73). They use the Batchelor's expression for the mobility and the
virial form for the osmotic pressure (69), such that for the case when the interaction
potential contains in addition to a hard core part a purely attractive interaction
the mobility \( b_{11}\left( \Phi \right)  \) increases and both, the compressibility
\( \left( \frac{\partial p_{B}}{\partial \rho _{B}}\right) _{T} \) and the
collective diffusion \( D_{c} \) decrease. The opposite effect occurs when
a purely repulsive interaction appears along with the hard repulsive part. These
results have been corroborated by experiments \cite{russel3}.

We would like to notice on the fact that although our study has been performed
for interacting Brownian particles the obtained collective diffusion coefficient
(73) could be used to study sedimentation processes \cite{russel}. Moreover,
to interpret recent experiments of sedimenting particles \cite{segre}, the
functional form of the osmotic pressure \( p_{B}\left( \Phi \right)  \) for
hard sphere Brownian particles is particularly useful. Our study focused on
continuous pair interactions, we then confirm the validity of the relation between
collective diffusion, mobility and compressibility (eq. 73) for any pair effective
interaction.

\section{\protect\smallskip Concluding Remarks}

In this paper we have analyzed the dynamics of a set of interacting Brownian
particles by using the method of nonequilibrium thermodynamics. Our approach
starts from a previous treatment of simultaneous Brownian motion of colloidal
particles, for which the Fokker-Planck equation, describing the evolution of
the \( N \)-particle distribution function, was derived in the framework of
mesoscopic nonequilibrium thermodynamics. The evolution equations for the \( s \)-particle
distribution functions are obtained from the continuity equation in phase space.
The currents ocurring in those equations follow from mesoscopic nonequilibrium
thermodynamics. The first two equations are the one- and two-particle Fokker-Planck
equations.

The corresponding Fokker-Planck-like equations are analogous to the ones obtained
by Mazo \cite{mazo} and Piasecki \textsl{et al.} \cite{piasecki}, for the
case of hard-sphere macroparticles in a hard-sphere fluid with a multiple time
scale which permits the derivation of microscopic expressions for the friction
tensors between the macroparticles. The account of the discrete nature of the
solvent is very useful to describe, for example, the dynamics of nanocolloids
\cite{Heyes}.

\smallskip From the kinetic equations we have discussed the hydrodynamic regime.
The balance equations for the conserved quantities are derived from the equations
relating those quantities with the moments of the distributions. The momentum
balance permits the analysis of the diffusion regime emerging at long times.
In this particular level of description, we obtain a generalized Einstein relation
for the collective diffusion, mobility and compressibility of the Brownian particles,
where the excess chemical potential and the mobility stand for the fit parameters.\\

\noindent \pagebreak
\textbf{\large Acknowledgements}\\
 {\large \par{}}{\large \par}

M.M. and L.R-S acknowlegde partial financial support from CGIyEA (UAEM\'{e}x)
\ and CONACyT (M\'{e}xico) by the projects J32094-E and J33080-E respectively.
J.M. Rub\'{\i} acknowledges financial support from DGICyT of the Spanish Goverment,
under grant PB98-1258.\smallskip  The authors acknowledge Dr. I. Pagonabarraga
for a critical reading of the manuscript.

\end{document}